\documentclass[conference,a4paper]{IEEEtran}
\usepackage{xcolor}
\usepackage{balance}
\usepackage{amsmath}
\usepackage{amssymb}
\ifCLASSINFOpdf
  \usepackage[pdftex]{graphicx}
\else
  \usepackage[dvips]{graphicx}
\fi
\ifCLASSOPTIONcompsoc
 \usepackage[caption=false,font=normalsize,labelfont=sf,textfont=sf]{subfig}
\else
 \usepackage[caption=false,font=footnotesize]{subfig}
\fi
\begin{document}
%
\title{Interpolation of impedance matrices for varying quasi-periodic boundary conditions in 2D periodic Method of Moments}

\author{\IEEEauthorblockN{
Denis Tihon\IEEEauthorrefmark{1},   
Christophe Craeye\IEEEauthorrefmark{2},   
Nilufer Ozdemir\IEEEauthorrefmark{3},    
Stafford Withington\IEEEauthorrefmark{1}      
}                                     
\IEEEauthorblockA{\IEEEauthorrefmark{1}
Cavendish Laboratory, University of Cambridge, Cambridge, UK.}
\IEEEauthorblockA{\IEEEauthorrefmark{2}
ICTEAM Institute, Universit\'{e} catholique de Louvain, Louvain-la-Neuve, Belgium.}
\IEEEauthorblockA{\IEEEauthorrefmark{3}
Scientific and Technological Research Council of Turkey, Space Technologies Research Institute, Turkey}
 \IEEEauthorblockA{Corresponding author: dt501@cam.ac.uk }
}



\maketitle

\begin{abstract}
Periodic structures can be simulated using the periodic Method of Moments. The quasi-periodicity, i.e. periodicity within a linear phase-shift, is implemented through the use of the periodic Green's function. In this paper, we propose a technique to interpolate the impedance matrix for varying phase-shifts. To improve the accuracy, the contribution of the dominant Floquet modes and a term corresponding to a linear phase-shift are first extracted. The technique is applied to planar geometries, but can be extended to non-planar configurations.
{\color{blue} This paper has been published in the proceedings of the 15th European Conference on Antennas and Propagation (EuCAP 2021). The original version of the paper is available at https://doi.org/10.23919/EuCAP51087.2021.9411301.
}\end{abstract}

\vskip0.5\baselineskip

%

\section{Introduction}
Electromagnetic periodic structures have received a lot of attention due to their peculiar properties and their relatively easy integration in complex designs. Among the applications are the design of Frequency-Selective Surfaces (FSS), reflect- and transmit-arrays, metasurfaces, metamaterials, etc. The efficient study of these materials requires suitable numerical tools.

Structures periodic in two directions can be simulated using the Method of Moments (MoM), a spectral method based on surface integral equations \cite{ref1}. If the excitation is quasi-periodic, \textit{i.e.} it shares the periodicity of the structure within a linear phase-shift, the response of the structure can be obtained from the study of a single unit cell. The periodicity is accounted for through the use of a modified Green's function that accounts for the periodicity of the structure. First, the impedance matrix of the structure is computed. Then, the resulting system of equation is solved to obtain the response of the structure to a given excitation. For unit cells of small or moderate size, most of the time is devoted to the computation of the periodic impedance matrix. Non-periodic sources can be handled using the Array Scanning Method (ASM), which consists of decomposing the non-periodic source into a superposition of quasi-periodic sources and solving each subproblem separately \cite{ref6}.

One drawback of the periodic Method of Moments is that the response of the structure must be computed for each frequency and relative phase-shift between consecutive unit cells. This problem becomes critical when using the ASM, for which a large number of phase shifts may be required to get accuracte results \cite{ref9}. One way to counter this drawback is to use interpolation techniques \cite{ref12, ref10, ref11}. The value of the impedance matrix is computed for few reference frequencies and/or phase shifts. Its value at different frequencies and/or phase shifts is then interpolated. Brute-force polynomial interpolation is inefficient due to the rapidly oscillating behaviour of the Green's function and the singularity arising from plane-waves with grazing incidence. To improve the convergence of the method, it was proposed to use a Model-Based Parameter Estimation (MBPE) technique \cite{ref12, ref11}. Based on the physics, a model is developed to describe the generic behaviour of the periodic impedance matrix entries. Then, the evolution of each entry of the impedance matrix is obtained by adjusting few coefficients involved in the model. In \cite{ref10}, the periodic impedance matrix for normally incident plane-waves is divided into two contributions: the resonant modes of the substrate and the rest. Then, the relative contribution of the different terms is estimated to fit the actual value of the periodic impedance matrix at few sampling frequencies. In \cite{ref11}, it is proposed to extend the technique to oblique incidence by removing an additional phase term. In this way, frequency interpolation of the periodic impedance matrix can be carried out for non-vanishing phase shifts between consecutive unit cells. However, these studies concentrate on the interpolation of the periodic impedance matrix vs. frequency. It seems that the interpolation with respect to the phase shift has received little attention so far.

In this paper, we propose a method to interpolate the periodic impedance matrix for different phase shifts between consecutive unit cells. First, the contribution of the leading Floquet modes is removed from the impedance matrix. Then, an additional phase term is extracted. It is shown that the resulting function is smooth with respect to the phase shift between consecutive unit cells. The contribution of the dominant Floquet modes is evaluated numerically instead of being fitted, so that very few sampling points are required for the interpolation. It is in contrast with MBPE techniques for which at least one sampling point per coefficient is required. It is also shown that removing the contribution of the main Floquet modes breaks the periodicity of the impedance matrix, so that one impedance matrix can be used to obtain several sampling points.

The paper is organized as follows. In Section II, the working principle of the periodic MoM is briefly described. Then, in Section III, the behaviour of the periodic impedance matrix is studied and the interpolation technique is described. Last, in Section IV, the interpolation technique is validated through few examples. 

\section{The Periodic Method of Moments}
Surface Integral Equation based numerical methods are relying on the surface equivalence principle. The response of a structure to incident fields is modeled using equivalent currents. The fields generated by these currents correspond to the fields scattered by the structure. The amplitude of the equivalent currents for a given excitation can be determined by enforcing boundary conditions across each interface separating different media. Using the Method of Moments, the unknown currents are discretized using a predefined set of basis functions (BF) and the boundary conditions are imposed on average along a predefined set of testing functions (TF) \cite{ref8}. First, the impact of each BF on the boundary conditions along each TF is computed and stored in the so-called impedance matrix. Then, the amplitude of the currents is found by solving the resulting system of equations:
\newcommand{\mat}[1]{\underline{\underline{#1}}}
\newcommand{\vect}[1]{\mathbf{#1}}
\begin{equation}
\mat{Z} \, \vect{x} = - \vect{b},
\end{equation}
with $\mat{Z}$ the impedance matrix, $\vect{x}$ a vector containing the unknown coefficients used to expand the equivalent currents using the BF and $\vect{b}$ the impact of the incident fields on the boundary conditions.
One natural set of boundary conditions that can be used is the continuity of the tangential electric and magnetic fields, leading to the Poggio-Miller-Chang-Harrington-Wu-Tsai (PMCHWT) formulation. Using this formulation, the impedance matrix corresponds to the sum of the fields generated by the BF along the TF through the media on both sides of the interface.

This method can be extended to periodic structures provided that the fields radiated by replicas of the BFs are included when computing the impedance matrix \cite{ref1}. In that case, one obtains the following system of equations
\newcommand{\Zper}{\mat{\tilde{Z}}}
\begin{equation}
\Zper(\boldsymbol{\varphi}) \, \vect{x} = -\vect{b}(\boldsymbol{\varphi})
\end{equation}
with $\Zper$ the periodic impedance matrix and $\boldsymbol{\varphi}$ the phase-shift between consecutive unit cells in the two directions of periodicity. 

One way to compute the periodic impedance matrix is to first compute the the periodic Green's function and convolve it spatially with the BFs and integrate the result along the TF. In that case, several methods can be used to accelerate the computation of the periodic Green's function (see \textit{e.g.} \cite{ref2, ref3, ref4, ref5}). Another possibility is to compute directly the impedance matrix in the spectral domain. In that case, each entry of the impedance matrix can be expressed as an infinite series of spectral term, each term corresponding to one Floquet mode of the structure \cite{ref14}.

\section{Interpolation scheme}
\newcommand{\dir}[1]{\hat{\mathbf{#1}}}
We consider a homogeneous medium of permittivity and permeability $\varepsilon$ and $\mu$. In this medium, a quasi-periodic set of BF of period $d_x$ and $d_y$ and phase-shifts $\varphi_x$ and $\varphi_y$ in the $\dir{x}$ and $\dir{y}$ directions emit electric ($E$) and magnetic ($H$) fields, which are tested with the TFs. We define the direction $\dir{z} = \dir{x}\times \dir{y}$. We consider that the TF is located either above or below the BF, \textit{i.e.} the sign of $\dir{z} \cdot (\vect{r}' - \vect{r})$ is identical for any pair of points $\vect{r}'$ and $\vect{r}$ on the BF and TF, respectively. This condition is always met in planar geometries, so that we will restrain ourselves to this particular case for the rest of the paper. Then, using the notations of \cite{ref9}, the corresponding entry of the periodic impedance matrix reads 
\newcommand{\sumInf}[1]{\sum_{#1=-\infty}^\infty}
\newcommand{\FTBe}{\tilde{f}_{B,e} (\mathbf{k}_{pq})}
\newcommand{\FTBm}{\tilde{f}_{B,m} (\mathbf{k}_{pq})}
\newcommand{\FTTe}{\tilde{f}_{T,e} (-\mathbf{k}_{pq})}
\newcommand{\FTTm}{\tilde{f}_{T,m} (-\mathbf{k}_{pq})}
\begin{align}
\tilde{Z}^{EJ}&(\boldsymbol{\varphi}) = 
\dfrac{\eta k}{2 d_x d_y}
\sumInf{p} \sumInf{q} \dfrac{1}{\gamma_{pq}}
\label{eq:01}
\\
&  \hspace{-1cm} \times
\Big( \FTBe \FTTe  + \FTBm \FTTm \Big)
\nonumber
\\
\tilde{Z}^{EM}&(\boldsymbol{\varphi}) = 
\dfrac{ k}{2 d_x d_y}
\sumInf{p} \sumInf{q}
\dfrac{1}{\gamma_{pq}}
\label{eq:02} 
\\
& \hspace{-1cm} \times
\Big( \FTBe \FTTm - \FTBm \FTTe \Big) \nonumber
\end{align}
with $\boldsymbol{\varphi} = (\varphi_x, \varphi_y)$, $\eta = \sqrt{\mu/\varepsilon}$ the impedance of the medium through which the interactions are computed, $k = \omega \sqrt{\varepsilon \mu}$ the wavenumber of the medium, $\omega$ the angular frequency, $\mathbf{k}_{pq} = (k_{x,p}, k_{y,q}, \pm \gamma_{pq})$ the wave-vector associated to Floquet harmonic $(p,q)$, $\tilde{f}_{B/T,e/m}$ the Fourier transform of the $\dir{e}$ or $\dir{m}$ component of the BF ($B$) and TF ($T$), with:
\begin{align}
\dir{e}(\mathbf{k}_{pq}) 
&= 
\dfrac{1}{k_{t,pq}} \big(-k_{x,p}, k_{y,q}, 0 \big) 
\\
\dir{m}(\mathbf{k}_{pq}) 
&=
- \dfrac{\gamma_{pq} }{k k_{t,pq}} \bigg(\pm k_{x,p}, \pm k_{y,q}, - \dfrac{k_{t,pq}^2}{\gamma_{pq}} \bigg)
\\
\mathbf{k}_{t,pq} &= \big(k_{x,p}, k_{y,q}, 0 \big) \\
\gamma_{pq} &= \sqrt{k^2-k_{t,pq}^2} \\
k_{t,pq} &= \sqrt{\mathbf{k}_{t,pq} \cdot \mathbf{k}_{t,pq}} \\
k_{x,p} &= \dfrac{2 \pi p + \varphi_x}{d_x} \label{eq:08}\\
k_{y,q} &= \dfrac{2 \pi q + \varphi_y}{d_y} \label{eq:09}
\end{align}
Note that where $\pm$ is indicated, the $+$ sign is used if the TF is located above the BF, and the $-$ sign otherwise. Similarly, note that $b = \sqrt{a}$ is defined such that its imaginary part is negative and, if $b$ is real, it is positive.

From \eqref{eq:01} and \eqref{eq:02}, the dependence of the impedance matrix on $\boldsymbol{\varphi}$ can be analyzed. Looking at each term of the sum separately and factoring out the phase term due to the spatial shift between the BF and TF, one can highlight four different factors: 
\newcommand{\FTBemo}{\tilde{f}_{B,e/m}^0 (\mathbf{k}_{pq})}
\newcommand{\FTTemo}{\tilde{f}_{T,e/m}^0 (-\mathbf{k}_{pq})}
\newcommand{\rb}{\mathbf{r}_B^0}
\newcommand{\rt}{\mathbf{r}_T^0}
\newcommand{\kpq}{\mathbf{k}_{pq}}
\begin{equation}
\label{eq:03}
\dfrac{1}{\gamma_{pq}}
\end{equation}
\begin{equation}
\label{eq:04}
\FTBemo \FTTemo 
\end{equation}
\begin{equation}
\label{eq:05}
\exp\bigg(j \mathbf{k}_{t,pq} \cdot (\rb-\rt)\bigg)
\end{equation}
\begin{equation}
\label{eq:06}
\exp\bigg(\pm j \gamma_{pq} \dir{z} \cdot (\rb-\rt)\bigg)
\end{equation}
with $\FTBemo$ and $\FTTemo$ the Fourier transforms of the BF and TF when they are centered at the origin and $\rb$ and $\rt$ the position of the BF's and TF's actual centers with respect to the origin.

We now look at each factor separately.
\subsubsection*{Factor \eqref{eq:03}}
For $k_{t,pq} \simeq k$, this factor varies rapidly vs. $\varphi_x$ and $\varphi_y$ that may be hard to interpolate. However, for $k_{t,pq} \gg k$, the variation is slow with respect to $\boldsymbol{\varphi}$ and thus the interpolation is expected to work well.
\subsubsection*{Factor \eqref{eq:04}}
The BF and TF are generally well-behaving functions, so their Fourier transform does not exhibit any sharp feature. Moreover, the BF and TF being usually much smaller than the unit cell, their Fourier transform is expected to vary very slowly with $\boldsymbol{\varphi}$.
\subsubsection*{Factor \eqref{eq:05}}
The variation of this factor with respect to $\boldsymbol{\varphi}$ does not depend on indices $(p,q)$ of the Floquet mode considered and can be factored out as 
\begin{equation}
\label{eq:07}
\exp\bigg(j \Big(\dfrac{\varphi_x}{d_x} \dir{x} + \dfrac{\varphi_y}{d_y} \dir{y}\Big) \cdot (\rb-\rt)\bigg).
\end{equation}
\subsubsection*{Factor \eqref{eq:06}}
Four different cases can be highlighted, depending on the distance $\Delta z$ between the BF and the TF in the $\dir{z}$ direction and depending on $k_{t,pq}$, the amplitude of the transverse wave-vector. On one hand, if $\Delta z \gg 0$, variations of the factor are rapid if $k_{t,pq} \lesssim k_0$. However, if $k_{t,pq} \gg k_0$, \eqref{eq:06} becomes negligible and the contribution of the corresponding terms to series \eqref{eq:01} and \eqref{eq:02} is negligible. On the other hand, if $\Delta z \simeq 0$, variation with respect to $\gamma_{pq}$ and thus $k_{t,pq}$ is slow. It should be noticed that for $k_{t,pq} = k$, despite the relatively slow variation, the function is not analytic. Thus, interpolation is not expected to provide accurate results when crossing this limit.

Summarizing all these observations, the high-order $(p,q)$ terms, and thus their sum, are expected to be easy to interpolate. Thus the interpolation method is the following:
\begin{enumerate}
\item Compute the periodic impedance matrix for few phase shifts.
\item Remove the contribution of the dominant Floquet modes ($|p|, |q| \leq N$).
\item Remove the phase term of \eqref{eq:07}.
\item Interpolate the remaining function for the phase shift of interest using a polynomial interpolation technique.
\item Re-introduce the phase term of \eqref{eq:07}.
\item Re-add the contribution of the dominant Floquet modes.
\end{enumerate}

\begin{figure}
\center
\includegraphics[width = 6 cm]{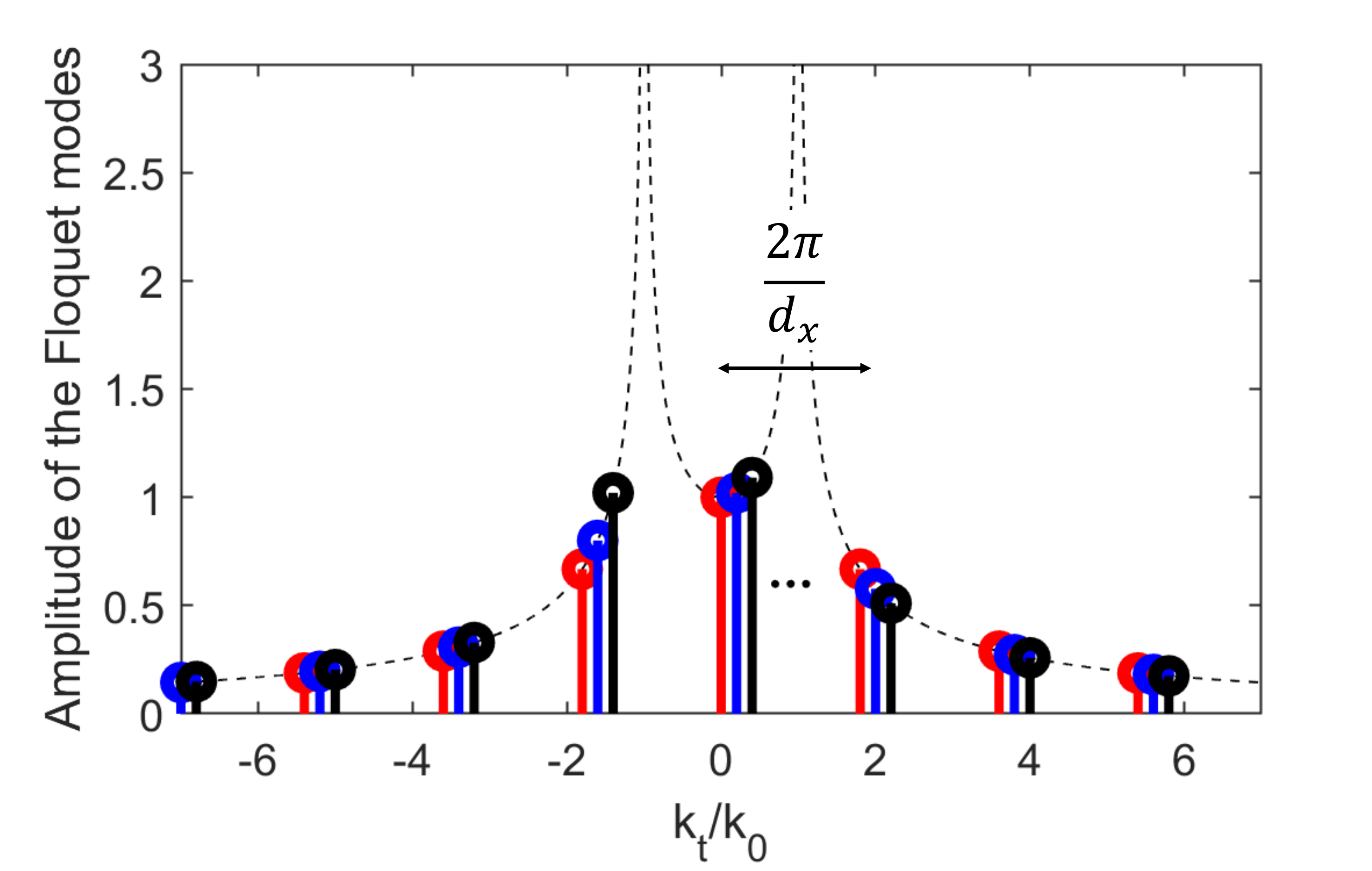} \\
\includegraphics[width = 7cm, trim={1cm 0 1.2cm 0},clip]{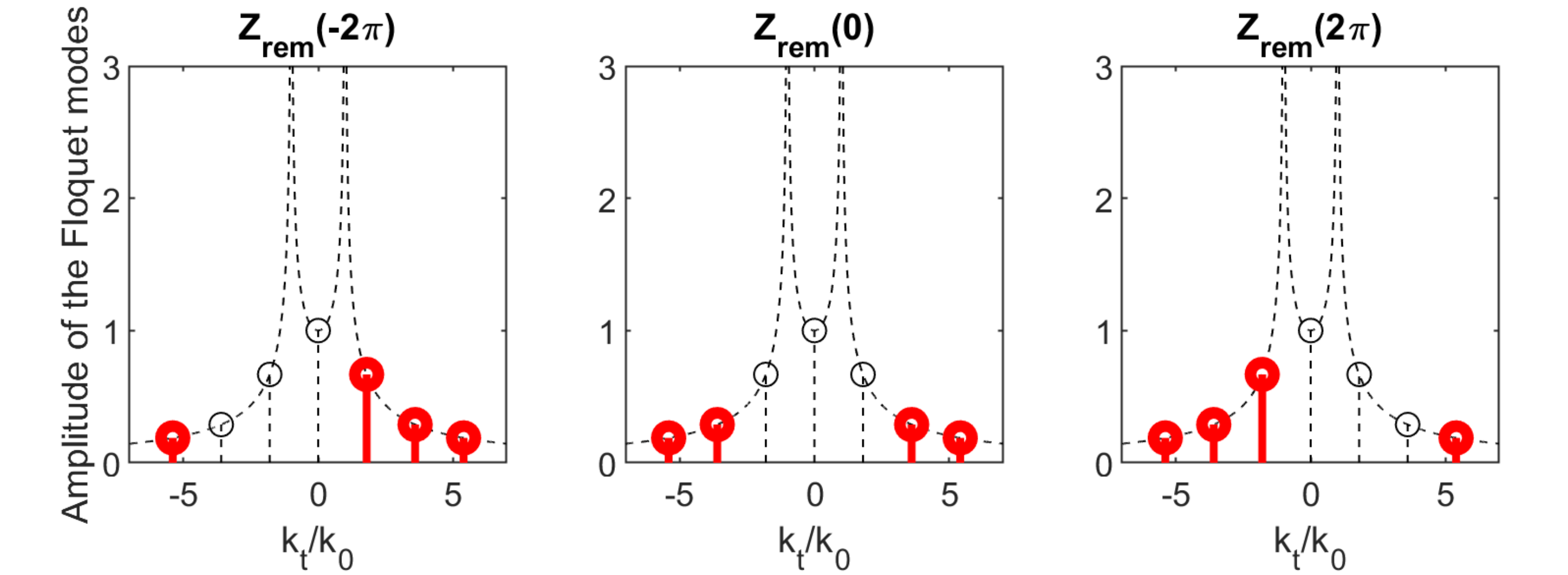}
\caption{Top graph: Qualitative illustration of the evolution of the Floquet modes of \eqref{eq:01} and \eqref{eq:02} with respect to the phase shift of the periodic impedance matrix. Different colors correspond to different phase shifts. Bottom graphs: Floquet modes included in the interpolant $Z_\text{rem}(\varphi)$ for $\varphi = -2\pi, 0, 2\pi$. The black dashed Floquet modes are not accounted for in order to smoothen the interpolant (Floquet modes extraction). Through the extraction of these modes, the periodicity of the series of \eqref{eq:01} and \eqref{eq:02} is broken.}
\label{fig:02}
\end{figure}

One interesting feature of the technique is that $\Zper$ is periodic with respect to $\varphi_x$ and $\varphi_y$, so that 
\begin{equation}
\Zper(\varphi_x, \varphi_y) = \Zper(\varphi_x + 2\pi, \varphi_y) = \Zper(\varphi_x, \varphi_y +2\pi).
\end{equation}
However, the Floquet modes that are removed do depend on the phase shift considered since, according to \eqref{eq:08} and \eqref{eq:09}, we have
\begin{equation}
k_{x,p}(\varphi_x +2\pi) = k_{x,p+1}(\varphi_x)
\end{equation}
\begin{equation}
k_{y,q}(\varphi_y + 2\pi) = k_{y,q+1}(\varphi_y)
\end{equation}
It means that from a single periodic impedance matrix, one can evaluate the value of the interpolant at several different locations. Let's take a simplified 1D example, as illustrated in Fig. \ref{fig:02}. On the top graph, one can see the evolution of the Floquet modes of the series \eqref{eq:01} and \eqref{eq:02} with the phase-shift $\varphi$.  Clearly, if $\varphi$ is increased by $2\pi$, the Floquet modes will be translated by one single period. Hence, when summing the contribution of the Floquet modes, one obtains the same value for $\varphi$ and $\varphi +2\pi$. However, after removing the contribution of the $2N+1$ dominant Floquet modes, the remaining terms in series \eqref{eq:01} and \eqref{eq:02} will be different for different phase shifts $\varphi$ (cf. bottom graphs). The resulting interpolant $Z_\text{rem}(\varphi)$ will be non-periodic. Thus, from a single periodic impedance matrix, by choosing properly the Floquet modes that are being removed, one can determine the value of the interpolant at several different locations.

For example, considering the original 2D periodic scenario, the periodic impedance matrix $\Zper(\boldsymbol{\varphi} = (0,0))$ can be used to estimate the value of the interpolant at points $(-2\pi, -2\pi)$, $(-2\pi, 0)$, $(-2\pi, 2\pi)$, $(0, -2\pi)$, $(0, 0)$, $(0, 2\pi)$, $(2\pi, -2\pi)$, $(2\pi, 0)$ and $(2\pi, 2\pi)$. For each of these sampling points, a different set of Floquet harmonics will be extracted from the periodic impedance matrix and thus the value of the interpolant will be different.

\section{Numerical results}
\begin{figure}
\center
\includegraphics[width = 8cm]{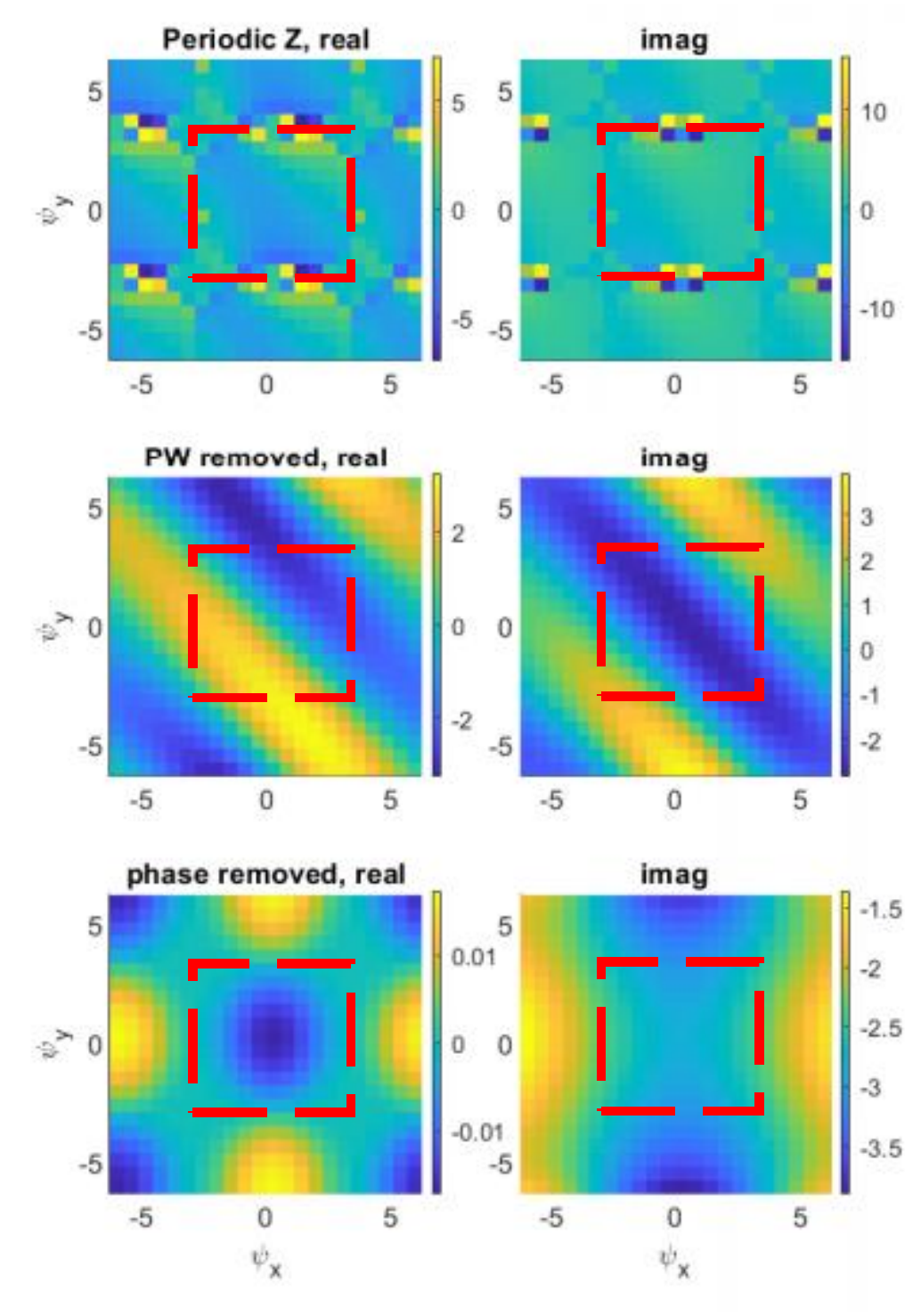}
\caption{Evolution of the periodic interaction with the contribution of all the Floquet modes (top graphs), without the contribution of the 9 dominant Floquet modes (middle graphs) and after removing the remaining phase term (bottom graphs). Both the real (LHS) and imaginary (RHS) parts of the functions are displayed. The red square delineates the zone of phase-shifts going from $-\pi$ to $\pi$.}
\label{fig:01}
\end{figure}

In order to illustrate the improvement of the interpolant with each successive transformation, we consider the interaction between two identical co-planar rooftop basis functions aligned in the $\dir{x}$ direction.  Each half of the rooftop is corresponding to a square of size $\lambda/18$, $\lambda$ being the wavelength. The periodicity in the $\dir{x}$ and $\dir{y}$ directions is $\lambda/1.8$. The TF corresponds to the replica of the BF translated by half a unit cell in the $\dir{x}$ and $\dir{y}$ directions. The evolution of the periodic interaction with phase shift is illustrated in the top graphs of Fig. \ref{fig:01}. Graphs in the left-hand side illustrate the real part and graphs on the right-hand side illustrate the imaginary part of the periodic interaction. The middle graphs illustrate the value of the periodic impedance matrix after removing the contribution of the 9 dominant Floquet modes ($N=1$). The bottom graphs illustrate the value of the interpolant after removing both the contribution of the 9 dominant Floquet modes and the phase term of \eqref{eq:07}. It can be seen that, at each step, the function becomes much smoother and easier to interpolate. Similar results are found for different pairs of BF and TF. It is worth noting that the value of the interpolant is illustrated for phase-shifts varying from $-2\pi$ to $2\pi$. While one is generally interested in estimating the periodic impedance matrix for phase shifts ranging from $-\pi$ to $\pi$, this extended zone can be used to add sampling points that may improve the polynomial interpolation. 

In order to check the accuracy of the interpolation procedure, we meshed a whole plane of the unit cell using 200 rooftop TFs oriented along the $\dir{x}$ and $\dir{y}$ directions. The periodic interaction between a single BF and the 200 TF was computed for several different heights $\Delta z$ between the BF and the TF. For the interpolation, we used four different periodic impedance matrices: $\Zper(0,0)$, $\Zper(0, \pi)$, $\Zper(\pi,0)$ and $\Zper(\pi, \pi)$. From these impedance matrices, the interpolant has been evaluated on 25 sampling points using the described procedure with $N=1$ (9 Floquet modes extracted), the coordinates of the sampling points corresponding to $\varphi_x, \varphi_y \in \{-2\pi, -\pi, 0, \pi, 2\pi\}$. Then, from these sampling points, the interpolant was approximated using a polynomial of order 4. Due to the high number of sampling points, the coefficients of the polynomial were obtained by solving an overconstrained system of equations, the relative weight attributed to the most distant points being a hundred times smaller than the weight attributed to the nine central points. 

Periodic impedance matrices were computed using the method of \cite{ref15}. Since only one BF is used, the resulting impedance ``matrices" correspond to vectors. The error is estimated as the norm of the difference between the interpolated impedance matrix and the computed one, relative to the norm of the computed one:
\begin{equation}
e(\boldsymbol{\varphi}) = \dfrac{\|\Zper_{interp}-\Zper\|_2}{\|\Zper\|_2}
\end{equation}
with $\Zper_{interp}$ the periodic impedance matrix obtained using the interpolation technique and $\|\cdot \|$ the $L2$ norm. For any distance and phase-shift, the maximum error was found to be approximately $0.2\%$.

We compared the times required to prepare the interpolation procedure and apply it. To do so, we considered the electric and magnetic fields generated by the plane on itself (200x400 interactions). In \cite{ref15}, the periodic Green's function is first tabulated and then integrated over the BF and TF. For each phase-shift the tabulation of the periodic Green's function required 20" and the computation of the impedance matrix required 6". Thus computing the four periodic impedance matrices required 1'45". Then, as explained earlier, the four periodic impedance matrices have been used to estimate the value of the interpolant at 25 different points. The estimation of the interpolant took about 10". Last, to interpolate the value of the matrix for a given phase-shift, it took about 0.5". The duration of the different steps is summarized in Table \ref{tab:01}. It should be emphasized that the computation of the reference impedance matrices and the extraction of the dominant Floquet modes only needs to be performed once.

\begin{table}
\center
\caption{Computation times involved in the interpolation technique (in seconds).}
\begin{tabular}{|c|c|c|}
\hline
Computation of four
&
Extraction of the dominant 
&
Interpolation of
\\
periodic impedance
&
Floquet modes and
&
one matrix
\\
matrices $\Zper$
&
linear phase shift
&
~
\\
\hline
105 & 10 & 0.5
\\
\hline
\end{tabular}
\label{tab:01}
\end{table}

Last, the interpolation technique has been used to compute the radiation pattern of the leaky-wave antenna proposed in \cite{sup1} and used in \cite{ref9} to validate the periodic MoM solver. The leaky wave antenna is made of a ground plane and a periodic array of patches located 19 mm above the ground plane. The patches are rectangular, with dimensions of 12.5 mm and 1 mm in the $\dir{x}$ and $\dir{y}$ directions. The periodicity of the patches is 13.5 mm in the $\dir{x}$ direction and 3 mm in the $\dir{y}$ direction. The structure is excited using a $\dir{x}$-directed dipole located 9.5 mm below the center of one patch. The radiation pattern of the structure in the $E$ and $H$ planes($x-z$ and $y-z$ planes, respectively) is studied at 9.5 GHz.

\begin{figure}
\center
\includegraphics[width = 7cm]{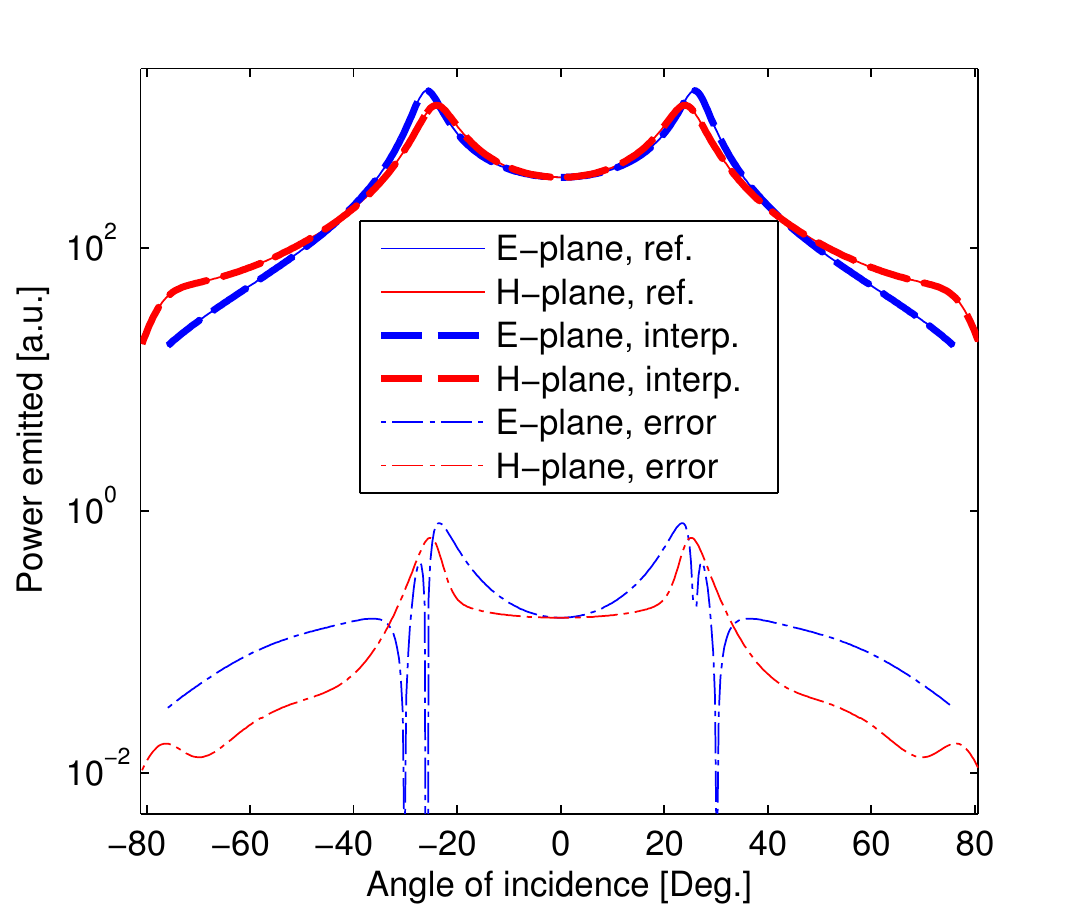}
\caption{Radiation pattern of the leaky-wave antenna described in \cite{sup1} and simulated in \cite{ref9}. Blue and red curves correspond to the radiation pattern in the E and H-planes, respectively. Continuous lines: reference radiation patterns for which each periodic impedance matrix was computed explicitly. Dashed lines: radiation patterns obtained using the interpolation technique. Dotted dashed lines: error due to the interpolation technique.}
\label{fig:03}
\end{figure}

The radiation pattern was computed twice, the results being displayed in Fig. \ref{fig:03}. In both cases, the code of \cite{ref9} was used to compute the periodic impedance matrices. The first radiation pattern was obtained by computing explicitly the periodic impedance matrix for each different phase shift. The second radiation pattern was obtained by computing the periodic impedance matrix for four different phase shifts and then interpolating the value of the periodic impedance matrix for the other phase shifts. The parameters used for the interpolation were the same as those used in the previous example. The radiation patterns obtained using both techniques and their difference are displayed in Fig. \ref{fig:03}. It can be seen that both radiation patterns are in perfect agreement and that the relative error introduced by the interpolation technique is about 0.1\%.

\section{Conclusion}
To conclude, we proposed a technique to efficiently interpolate the Method of Moments periodic impedance matrix vs. the phase shift between consecutive unit cells. To improve accuracy, the interpolant is smoothened by extracting the contribution of the dominant Floquet modes and a linear phase term. The efficiency of the method partly relies on the fact that a single periodic impedance matrix can be used to sample the interpolant at many different points. The accuracy of the method has been validated on numerical examples.

\section*{Acknowledgment}
This project has received funding from the European Union's Horizon 2020 research and innovation programme under the Marie Sk\l odowska-Curie grant agreement No. 842184.



%

\end{document}